\documentclass[aps,prl,floatfix,twocolumn]{revtex4}
\usepackage{amsmath,amssymb}
\usepackage{graphicx}
\usepackage{epsfig}
\usepackage{psfrag}
\newcommand{\bm}{\boldsymbol}

\begin{document}

\hsize\textwidth\columnwidth\hsize\csname@twocolumnfalse\endcsname

\title{Giant Magneto-optical Kerr Effect and Universal Faraday Effect
  in \\
Thin-film Topological Insulators}

\author{Wang-Kong Tse}
\author{A.~H. MacDonald}
\affiliation{Department of Physics, University of Texas, Austin, Texas 78712, USA}

\begin{abstract}
Topological insulators can exhibit strong magnetoelectric effects when their time-reversal symmetry is broken.  
In this Letter we consider the magneto-optical Kerr and Faraday effects of a topological 
insulator thin film weakly exchange-coupled to a ferromagnet. 
We find that its Faraday rotation has a universal value 
at low-frequencies, $\theta_{\mathrm{F}} = \mathrm{tan}^{-1}\,\alpha$ where 
$\alpha$ is the vacuum fine structure constant, and that 
it has a giant Kerr rotation $\theta_{\mathrm{K}} = \pi/2$.  These properties follow from 
a delicate interplay between thin-film cavity confinement and the surface Hall conductivity of a
topological insulator's helical quasiparticles. 
\end{abstract}

\maketitle
\newpage

When light scatters from a magnetic material, angular momentum is transferred to reflected (Kerr effect) and 
transmitted (Faraday effect) waves and induces polarization plane
rotations. 
The Kerr and Faraday magneto-optical effects allow light to directly probe time-reversal symmetry (TRS) breaking in a solid. 
Recently, topological insulators (TI) have aroused tremendous interest because of novel physics arising from the presence
of topologically dictated, gapless, Dirac-like surface states that respond singularly to disturbances which break TRS. 
Topological insulators can occur \cite{FuKaneMele,MooreBalents,Roy,SCZhangRev} when strong spin-orbit coupling drives band 
inversions. Recent angle-resolved photoemission spectroscopy (ARPES) experiments \cite{Hsieh1,Hsieh2,Noh,Nishide,BiSe_Xia,Hsieh4,Chen} have established that a number of compounds involving elements toward the lower right of the periodic table 
({\em e.g.} Bi$_x$Sb$_{1-x}$ and Bi$_2$Te$_3$, Bi$_2$Se$_3$) are topological insulators. 
When TRS is weakly broken, a gap $\Delta$ is induced at the Dirac points of a topological insulator 
and the surface states exhibit strong magneto-electric effects \cite{SCZhangRev,Vanderbilt}. 

So far, most theoretical studies of TI's have focused on the surface properties of a semi-infinite sample. 
The thin film geometry considered here is of particular interest and importance because any application based on the unique 
electronic properties of these
materials is likely to employ them in thin-film form. In this paper, we develop a theory of 
the magneto-optical properties of TI thin films.  We find that in the low frequency 
($\omega \ll \Delta, E_{\mathrm{g}}$ where $E_{\mathrm{g}}$ is the TI's bulk gap) limit 
the Faraday rotation is \textit{universal}, 
dependent only on the vacuum fine structure constant $\alpha = e^2/\hbar c = 1/137$. 
Our theory also predicts a low-frequency giant Kerr rotation $\theta_{\mathrm{K}} \simeq \pi/2$,
corresponding to a \textit{full-quarter} rotation of the polarization plane of light reflected from a TI thin film. 

The most important physical consequence of topologically entwined bulk bands is the 
appearance of surface states that must be gapless unless time-reveral symmetry is broken. 
A thin-film TI is characterized by metallic top and bottom surfaces 
that carry helical quasiparticles defined by a Dirac-like Hamiltonian\cite{HZhang1} $H
= (-1)^s v\bm{\tau}\cdot\bm{k}+\Delta\tau_z$, where $\bm{k} =
(k_x,k_y)$, $\bm{\tau} = (\tau_x,\tau_y,\tau_z)$ are 
Pauli matrices that act on the spin degrees of freedom, and $s = \pm 1$ labels the top ($+1$) and bottom ($-1$) surfaces. 
The first term in the Hamiltonian captures the gapless Dirac-cone energy spectrum of the surface states.
The influence of a TRS-breaking perturbation, when present, is captured by the {\em mass} term $\Delta\tau_z$, which opens up a gap at the Dirac point. 
As we discuss later, the mass term can be generated by Zeeman coupling, or by 
coating the TI surface with an insulating magnetic material \cite{SCZhangRev,Hsieh3} that is exchange-coupled to 
the TI surface states. We consider the TI film to be thick enough so that there is negligible
tunneling between the top and bottom surfaces. 
%

\textit{Electromagnetic response of a topological metallic surface} --- We begin by considering an electromagnetic wave normally incident on a topological metallic surface. The wave travels along the $z$ axis from a dielectric medium labeled by $i$, impinges on the topological surface at $z = a_i$, scatters, and then propagates into another dielectric medium labeled by $j$. The impinging electromagnetic field induces currents on the top and bottom surfaces of the TI with surface current densities $\tilde{J} = \bar{\sigma}\tilde{E}$.
Here tilde accents denote vectors, {\em e.g.} $\tilde{J} =
[J_x\,\,\,J_y]^{\mathrm{T}}$, and $\bar{\sigma}$ is the conductivity
tensor. 
%
%
We evaluate the electrical response of the TI surface
using a linearized collisionless quantum kinetic equation \cite{Rammer}:
\begin{equation}
\frac{\partial f_k^{(1)}}{\partial t}+e\bm{E}\cdot\frac{\partial f_k^{(0)}}{\partial \bm{k}}+i\left[v\bm{\tau}\cdot\bm{k}+\Delta\tau_z,f_k^{(1)}\right] = 0, \label{ke}
\end{equation}
where $f_k$ is a $2\times 2$ $k$-dependent matrix distribution function that 
encodes $\uparrow$ and $\downarrow$ spin-density contributions in its diagonal 
elements and transverse spin-density contributions (coherence between $\uparrow$ and $\downarrow$ spins) in its off-diagonal elements.
The superscripts denote order in an electric-field expansion. 
Eq.~(\ref{ke}) is valid when the Fermi level lies in a well-formed 
gap, the circumstance in which we will be most interested, and
otherwise for $\omega >\tau_{\mathrm{c}}^{-1}$ where $\tau_{\mathrm{c}}^{-1}$ is the collision
rate. The linear-response current is related to $f_k$ by 
$\bm{J} = e\,m \; \mathrm{Tr}\sum_k(\bm{j}_kf_k^{(1)})$, where
$\bm{j}_k = e\partial H/\partial\bm{k} = ev\bm{\tau}$ 
and $m$ is an odd integer equal to the number of surface Dirac cones. 
In the ensuing discussion, we focus on $m = 1$.  Results reported below can be adapted to 
TI's with more than one Dirac cone by letting $\bar{\sigma} \to m 
\bar{\sigma}$ and the surface carrier density $n_{\mathrm{s}} \to n_{\mathrm{s}}/m$.

When the TRS-breaking coupling parameter $\Delta$ is not zero, incident light induces both 
longitudinal and Hall current response on the TI surfaces.  
Recent experiments have demonstrated the feasibility of tuning the Fermi level $\mu_F$ of the topological surface by alkali atom deposition, 
molecular adsorption, or photo-doping \cite{TIdopingexp}.
We have therefore solved Eq.~(\ref{ke}) as a function of Fermi energy
position $\mu_F$ relative to the Dirac point, 
and used the result to evaluate 
the longitudinal conductivity $\sigma_{xx}(\omega) = \sigma_{xx}^{\mathcal{R}}+i\sigma_{xx}^{\mathcal{I}}$ and the 
Hall conductivity $\sigma_{xy}(\omega) =
\sigma_{xy}^{\mathcal{R}}+i\sigma_{xy}^{\mathcal{I}}$. 
In the following we express $\bar{\sigma}$ in $e^2/\hbar = \alpha \,
c$ units and set $c = 1$ except where specified. 
For the dissipative components of the optical conductivity we find that:  
\begin{eqnarray}
\sigma_{xx}^{\mathcal{R}} &=& \theta\left(\vert\mu_F\vert-\vert\Delta\vert\right)\delta(\omega)\left(\mu_F^2-\Delta^2\right)/4\vert\mu_F\vert \nonumber \\
&&+\left[{1}/{16}+(\Delta/2\omega)^2\right]\theta\left[\vert\omega\vert-2\,\mathrm{max}\left(\vert\mu_F\vert,\vert\Delta\vert\right)\right], \label{sxx} \nonumber \\
\sigma_{xy}^{\mathcal{I}} &=&
({\Delta}/4{\omega}) \;\theta\left[\vert\omega\vert-2\,\mathrm{max}\left(\vert\mu_F\vert,\vert\Delta\vert\right)\right], \label{sxy}
\end{eqnarray}
where $\theta(x)$ is the unit step function. The terms proportional to $\theta$ in $\sigma_{xx}^{\mathcal{R}}$ and $\sigma_{xy}^{\mathcal{I}}$ 
are due to interband absorption at frequencies $\omega$ above 
$2\mu_F$ when the Fermi energy value $\mu_F$ exceeds the gap. 
For the reactive components of the optical conductivity $\sigma_{xx}^{\mathcal{I}}$ and $\sigma_{xy}^{\mathcal{R}}$, which are due to off-shell virtual transitions, we find that: 
\begin{eqnarray}
\sigma_{xx}^{\mathcal{I}} &=&
\theta\left(\vert\mu_F\vert-\vert\Delta\vert\right)
(1/4\pi)\left(\mu_F^2-\Delta^2\right)/\omega\vert\mu_F\vert \nonumber \\
&&+(1/8\pi)\left\{-2(\Delta^2/\omega)\left\{\varepsilon_{\mathrm{c}}^{-1}-\left[\mathrm{max}\left(\vert\mu_F\vert,\vert\Delta\vert\right)\right]^{-1}\right\}\right. \nonumber \\
&&\left.+({1}/{2})\left[1+4({\Delta}/{\omega})^2\right]f(\omega)\right\} ,
\label{Isxx} \nonumber \\
\sigma_{xy}^{\mathcal{R}} &=& -(\Delta/4\pi\omega)f(\omega),
\label{Rsxy} 
\end{eqnarray}
where $f(\omega) =
\mathrm{ln}\left\vert(\omega+2\varepsilon_{\mathrm{c}})/(\omega-2\varepsilon_{\mathrm{c}})\right\vert
-
\mathrm{ln}\left\vert[\omega+2\,\mathrm{max}\left(\vert\mu_F\vert,\vert\Delta\vert\right)]/[\omega-2\,\mathrm{max}\left(\vert\mu_F\vert,\vert\Delta\vert\right)]\right\vert$,
and $\varepsilon_{\mathrm{c}}$ is the energy cut-off of the Dirac
Hamiltonian, which we associate with the separation between the Dirac
point and the closest bulk band.
(Henceforth we assume that $\Delta > 0$.  Results for $\Delta
< 0$ can be obtained by simply replacing $\Delta \to \vert\Delta\vert$
and inverting the sign of $\sigma_{xy}$.)
Eqs.~(\ref{sxx}) and Eqs.~(\ref{Rsxy}) agree with results obtained previously for  
the $\Delta=0$ \cite{Falkovsky} and $\omega=0$ \cite{Sinitsyn} limits of the Dirac model.

When combined with standard electrodynamic boundary conditions, the surface Hall conductivity in our theory  
captures the {\em axion field} contribution to the Maxwell Lagrangian \cite{SCZhangRev} 
and generalizes the effect to finite frequency and arbitrary TRS-breaking coupling strength $\Delta$. 
%
%
%
%
\begin{figure}[tbp]
\includegraphics[width=9cm,angle=0]{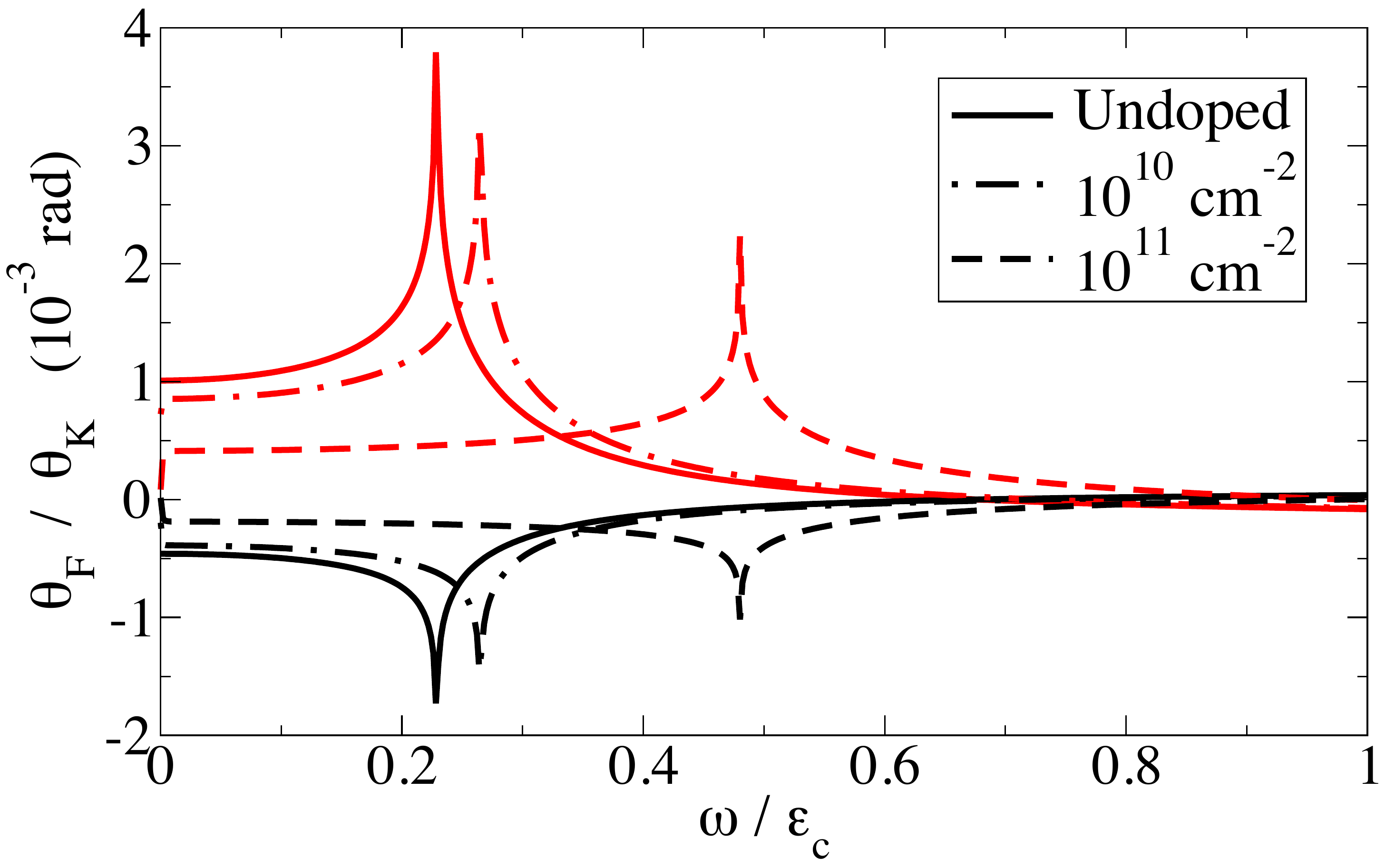} 
\caption{(Color online) Faraday and Kerr spectra
for a single vacuum/TI interface. $\theta_{\mathrm{F}}$ (red/grey) and
$\theta_{\mathrm{K}}$ (black) at three different carrier 
densities for $\Delta = 20\,\mathrm{meV}$ and the Bi$_2$Se$_3$ parameters $\epsilon = 29$, $\mu = 1$, $v = 5\times10^5\,\mathrm{ms}^{-1}$. 
We take $\varepsilon_{\mathrm{c}} = 0.175\,\mathrm{eV}$, half of the bulk band gap.  
} \label{fig2}
\end{figure}
%
%
%
By composing top and bottom surface scattering processes of a TI
thin film, we express the transmitted $\tilde{E}^{\mathrm{t}}$ and reflected
$\tilde{E}^{\mathrm{r}}$ components of the electric field as
%
\begin{eqnarray}
\tilde{E}^{\mathrm{t}} &=& \bar{t}_{\mathrm{B}}\left(\bm{1}-\bar{r}_{\mathrm{T}}'\bar{r}_{\mathrm{B}}\right)^{-1}\bar{t}_{\mathrm{T}} \, \tilde{E}^{0}, \label{trans} \\
\tilde{E}^{\mathrm{r}} &=& \left[\bar{r}_{\mathrm{T}}+\bar{t}_{\mathrm{T}}'\bar{r}_{\mathrm{B}}\left(\bm{1}-\bar{r}_{\mathrm{T}}'\bar{r}_{\mathrm{B}}\right)^{-1}\bar{t}_{\mathrm{T}}\right]\tilde{E}^{0}, \label{refl}
\end{eqnarray}
where $\tilde{E}^0$ is the incident field, $\bar{r},\bar{r^{'}}$ and
$\bar{t},\bar{t^{'}}$ are $2\times 2$ single-surface reflection and transmission
tensors, and the subscripts `T' and `B' label the top and bottom surfaces. Eqs.~(\ref{trans}) and (\ref{refl}) express the transmitted and reflected fields in
the form of geometric series resulting from Fabry-Perot-like repeated scattering at the top and bottom surfaces. 
%

\textit{Faraday and Kerr rotations} --- The Faraday and Kerr angles can be defined by $\theta_{\mathrm{F}} =
\left(\mathrm{arg}\{E_{+}^{t}\}-\mathrm{arg}\{E_{-}^{t}\}\right)/2$
and $\theta_{\mathrm{K}} =
\left(\mathrm{arg}\{E_{+}^{r}\}-\mathrm{arg}\{E_{-}^{r}\}\right)/2$ where ${E}_{\pm}^{r,t} = {E}_x ^{r,t}\pm i{E}_y ^{r,t}$ are the left-handed (+) and right-handed (-) polarized 
components of the outgoing electric fields, and the incoming electric field is linearly polarized.
For comparison purposes, we first report results for the case of single-interface transmission  
from vacuum to the bulk of a semi-infinite TI.     
We use Bi$_2$Se$_3$ which has a single Dirac cone and a large bulk band gap $E_{\mathrm{g}} \simeq 0.35\,\mathrm{eV}$
 \cite{BiSe_Xia} as an example. 
Fig.~\ref{fig2} shows results for three different surface carrier density values. 
We note that $\theta_{\mathrm{F}}$ and $\theta_{\mathrm{K}}$ are sharply peaked at the 
interband absorption threshold, most strongly so in the undoped
case.  In the low frequency ($\omega \ll \varepsilon_{\mathrm{c}} \sim E_{\mathrm{g}}$) limit of the 
undoped case
we have from Eqs.~(\ref{sxx})-(\ref{Rsxy}) that $\sigma_{xx}^{\mathcal{R}} = \sigma_{xx}^{\mathcal{I}} = 0$, $\sigma_{xy}^{\mathcal{I}} = 0$ and $\sigma_{xy}^{\mathcal{R}} =
(\alpha/4\pi)(1-\Delta/\varepsilon_{\mathrm{c}})$.
%
%
For $\Delta \ne 0$, both top and bottom TI surfaces exhibit 
dissipationless half-quantized Hall effects, similar to those 
found in graphene \cite{graphene_qhe}, but without the additional spin and 
valley degeneracies.  The half-quantized Hall effect \cite{half} is a particularly 
striking example of an intrinsic anomalous Hall effect \cite{sinova_RMP} 
contribution associated with momentum-space Berry phases \cite{niu_RMP}.  
It follows that 
\begin{equation}
\theta_{\mathrm{F}} = \mathrm{tan}^{-1}\left[\alpha
    (1-\Delta/\varepsilon_{\mathrm{c}})/\left(\sqrt{\epsilon_i/\mu_i}+\sqrt{\epsilon_j/\mu_j}\right)\right], \label{Fara_SL} 
\end{equation}
\begin{equation}
\theta_{\mathrm{K}} = \mathrm{tan}^{-1}\left[\frac{2\alpha\sqrt{\epsilon_i/\mu_i}(1-\Delta/\varepsilon_{\mathrm{c}})}{\sqrt{\epsilon_i/\mu_i}-\sqrt{\epsilon_j/\mu_j}-\alpha^2 (1-\Delta/\varepsilon_{\mathrm{c}})^2}\right]. 
\label{Kerr_SL}
\end{equation}
For $\Delta \ll \varepsilon_{\mathrm{c}}$, we obtain Eq.~(106) of Ref.~\cite{SCZhangRev} as a limiting case of Eq.~(\ref{Fara_SL}).
As expected $\theta_{\mathrm{F}}$ and $\theta_{\mathrm{K}}$ are dependent on the specific material properties of the TI; for the Bi$_2$X$_3$ ($\mathrm{X} = \mathrm{Te, Se}$)
class of materials the magnetic permeability $\mu \approx 1$ but the dielectric constant $\epsilon \sim 30 -80$ is large, reducing both $\theta_{\mathrm{F}}$ and $\theta_{\mathrm{K}}$ to $\sim 10^{-3}\,\mathrm{rad}$.
\begin{figure}[tbp]
 \includegraphics[width=8.7cm,angle=0]{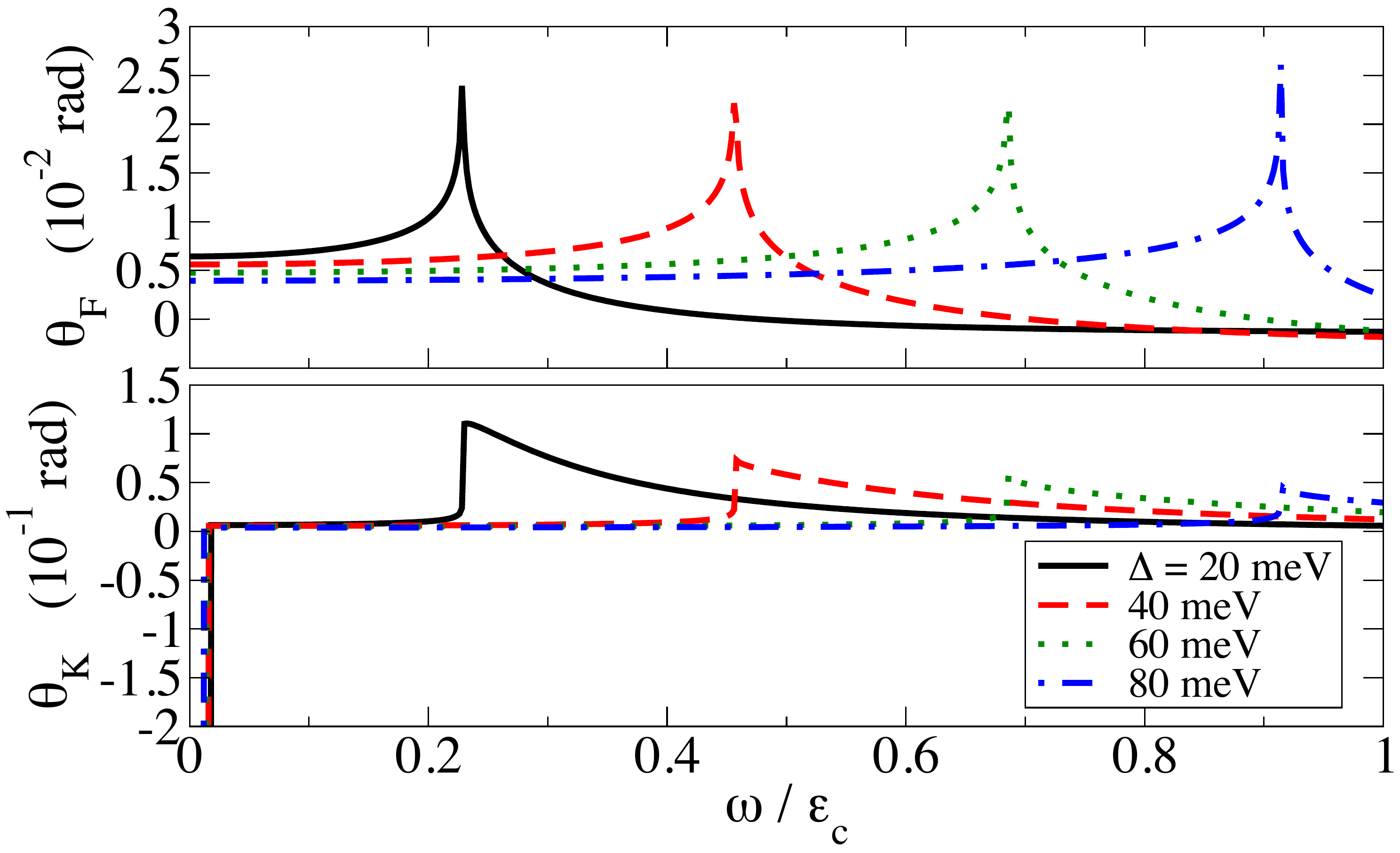} 
\caption{(Color online) Faraday and Kerr spectra for a TI thin film. $\theta_{\mathrm{F}}$ and $\theta_{\mathrm{K}}$ as a function of frequency at a series of $\Delta$ values
  for a $30\,\mathrm{nm}$-thick Bi$_2$Se$_3$ thin film with the Fermi
  level inside the Dirac gap on both top and bottom surfaces.  Notice that $\theta_{\mathrm{K}}$ values for $\omega \to 0$ lie well beyond the range of this figure.}
 \label{fig3}
\end{figure}

We now consider the thin film configuration and focus on the simplest and most 
interesting limit, that in which the Fermi levels of both surfaces lie
in the Dirac gap.
This limit can in principle be achieved experimentally by bulk hole doping \cite{Hsieh3}. 
Fig.~\ref{fig3} shows our results for $\theta_{\mathrm{F}}$ and $\theta_{\mathrm{K}}$, which are enhanced by over an order of magnitude
compared to the the single-interface case. 
The interband absorption features are also more prominent, with $\theta_{\mathrm{F}}$ showing a stronger peak and $\theta_{\mathrm{K}}$ exhibiting an abrupt jump at the 
interband threshold. The magnitude enhancement can be attributed to thin-film Fabry-Perot cavity confinement, in which multiple internal reflections enhance both rotations. 
In the low frequency regime $\omega \ll E_{\mathrm{g}}$, we have discovered 
universal features in both $\theta_{\mathrm{F}}$ and
$\theta_{\mathrm{K}}$ \cite{Remark}.  For the Faraday rotation 
we obtain from Eq.~(\ref{trans}) that 
the transmitted field $\tilde{E}^{\mathrm{t}} =
\{1+(4\pi\sigma_{xy}^{\mathcal{R}})^2\}^{-1}[1\, , \,4\pi\sigma_{xy}^{\mathcal{R}}]^{\mathrm{T}}$. 
It follows that the Faraday rotation $\theta_{\mathrm{F}} =
\mathrm{tan}^{-1}(4\pi\sigma_{xy}^{\mathcal{R}}) =
\mathrm{tan}^{-1}[\alpha(1-\Delta/\varepsilon_{\mathrm{c}})] \simeq \mathrm{tan}^{-1}\alpha$.
%
%
Provided that $\Delta \ll \varepsilon_{\mathrm{c}}$,
$\theta_{\mathrm{F}}$ is universal for a a TI thin film. 
This should be contrasted with the single-surface case (Eq.~(\ref{Fara_SL}) and Ref.~\cite{SCZhangRev}), for 
which $\theta_{\mathrm{F}}$ is explicitly  $\epsilon$ and $\mu$
dependent. The universal low-frequency limits of the Faraday and Kerr
angles 
are not altered \cite{Remark} by a substrate dielectric. 
\begin{figure}[tbp]
  \includegraphics[width = 8.8cm,angle=0]{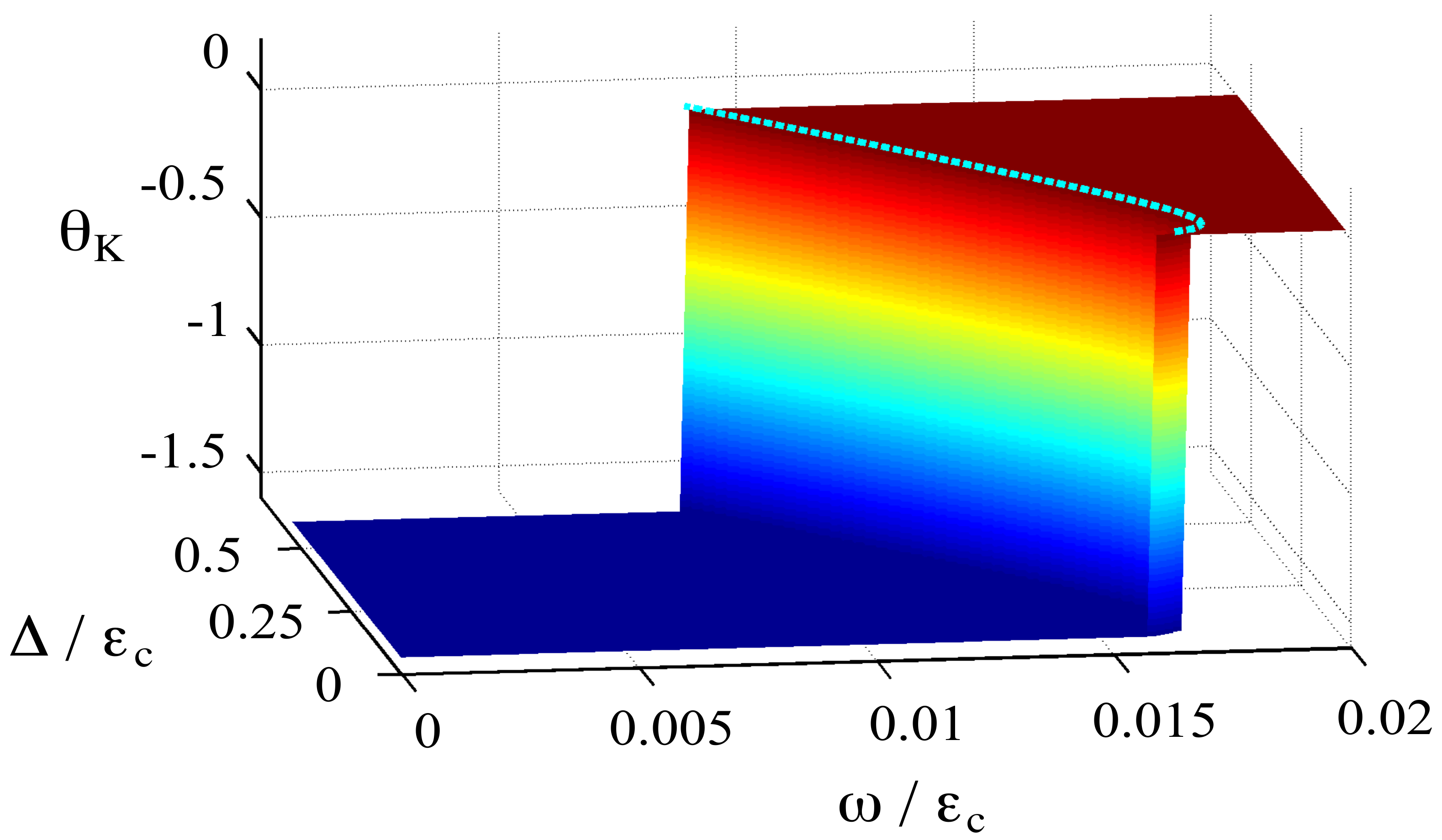}
\caption{(Color online) Low-frequency Kerr rotation. Kerr rotation as a function of frequency
  $\omega/\varepsilon_{\mathrm{c}}$ and gap
  $\Delta/\varepsilon_{\mathrm{c}}$, showing a ``waterfall'' feature
  with 
  $\theta_{\mathrm{K}}$ exhibiting a constant value $\simeq -1.56 =
  -\pi/2$ at low frequencies $\omega \lesssim
  0.02\varepsilon_{\mathrm{c}}$. Dashed line shows $\omega_{\mathrm{K}}$ from Eq.~(\ref{omegaK}). 
The approximate formula is in good agreement with numerical results. 
} \label{fig4}
\end{figure}
For the Kerr rotation we examine the reflected field
$\tilde{E}^{\mathrm{r}} = \{1+(4\pi\sigma_{xy}^{\mathcal{R}})^2\}^{-1}[-(4\pi\sigma_{xy}^{\mathcal{R}})^2\, ,\,4\pi\sigma_{xy}^{\mathcal{R}}]^{\mathrm{T}}$.
We find that for $\Delta>0$, $\theta_{\mathrm{K}}
=-\mathrm{tan}^{-1}(1/4\pi\sigma_{xy}^{\mathcal{R}}) \simeq -\pi/2,$ 
%
%
{\em i.e.} the polarization of the reflected light exhibits a
striking \textit{full-quarter} rotation relative to the incident
polarization direction. 
This result is numerically insensitive to 
the precise value of the gap as long as it is finite.  
This giant enhancement stems from two factors which combine to give an overall large effect: 
the topological nature of the surface transport for $\omega \ll E_{\mathrm{g}}$ and interference between the 
immediate reflection wave with waves that have been scattered off the bottom surface. 
First, the peculiarity of the Dirac helical quasiparticles 
gives rise to dissipationless Hall transport on the TI surface even when the Fermi level $\mu_F$
lies in the Dirac gap. The surface Hall response creates a splitting 
between the reflected left-handed and 
right-handed circularly polarized fields by $\pm
\sigma_{xy}^{\mathcal{R}} \sim \pm\alpha$ along the transverse
direction. 
The presence of the bottom topological surface is crucial since the first-order partial wave 
reflected from the top surface undergoes a relatively minor rotation $\sim\mathrm{tan}^{-1}\alpha$.
The higher-order partial waves from subsequent scattering with the bottom surface sum up geometrically to yield a contribution that interferes
destructively with the first-order partial wave along the incident
polarization plane, strongly suppressing it by a factor of $\sim
\sigma_{xy}^2 \propto \alpha^2$. As a result, the left-handed and
right-handed circularly polarized fields each acquire a phase angle 
$\sim \mathrm{tan}^{-1}(\alpha/\alpha^2) =
\mathrm{tan}^{-1}(1/\alpha) \simeq \pi/2$ in \textit{opposite}
directions. This leads to the large Kerr angle
$\vert\theta_{\mathrm{K}}\vert \simeq \pi/2$. 
%
%

Our numerical results also show that the giant enhancement of the Kerr
effect persists beyond the strictly {\em DC} limit, 
and that $\theta_{\mathrm{K}}$ remains close to $-\pi/2$ in a small but finite
frequency window around $\omega = 0$. In Fig.~\ref{fig4} we show an
enlarged plot of $\theta_{\mathrm{K}}$ in the vicinity of $\omega = 
0$. We also find that the enhancement is robust for a wide range of
gap values $\Delta$. For small frequencies, the reflected circularly
polarized components can be decomposed 
into separate leading-order contributions from the topological conducting surface and the thin
film dielectric, $E^r_{\pm} \simeq \pm i4\pi\sigma_{xy}^{\mathcal{R}}-i4\pi\sigma_{xx}^{\mathcal{I}}+i(\sqrt{\epsilon/\mu}-\sqrt{\mu/\epsilon})kd/2$, 
where $k = \omega\sqrt{\epsilon\mu}/c$ is the wave vector in the TI
and $d$ is the film thickness. 
The real components of $E^r_{\pm}$ are smaller by a factor $\sim
\alpha$ in this regime.
As frequency increases, the dielectric 
contribution to the imaginary part of $E^r_{\pm}$ 
eventually dominates so that the $\pm$ components have the same sign and 
$\theta_{K}$ rapidly falls to a small value. 
The frequency range for which giant Kerr angles occur 
is approximately given by 
\begin{equation}
\omega_{\mathrm{K}} \simeq 4\pi\sigma_{xy}^{\mathcal{R}}(0)/\left[\left(\epsilon-\mu\right)d/2c-4\pi{\sigma_{xx}^{\mathcal{I}}}'(0)\right].
\label{omegaK}
\end{equation}
This formula demonstrates 
that the frequency range over which 
$\vert\theta_{\mathrm{K}}\vert \simeq \pi/2$ can be widened 
by reducing the film thickness $d$ or the bulk 
dielectric constant $\epsilon$. 
For standard film thickness $10 - 100\,\mathrm{nm}$, we find that 
$\omega_{\mathrm{K}} \sim 5\,-\,70\,\mathrm{cm}^{-1}$ 
which translates to a wavelength $\lambda \sim
10^{-4}\,-\,10^{-3}\,\mathrm{m}$, 
implying that the predicted giant Kerr effect should
be measurable for frequencies up to the far infrared regime. 
\begin{figure}[tbp]
  \includegraphics[width=8cm,angle=0]{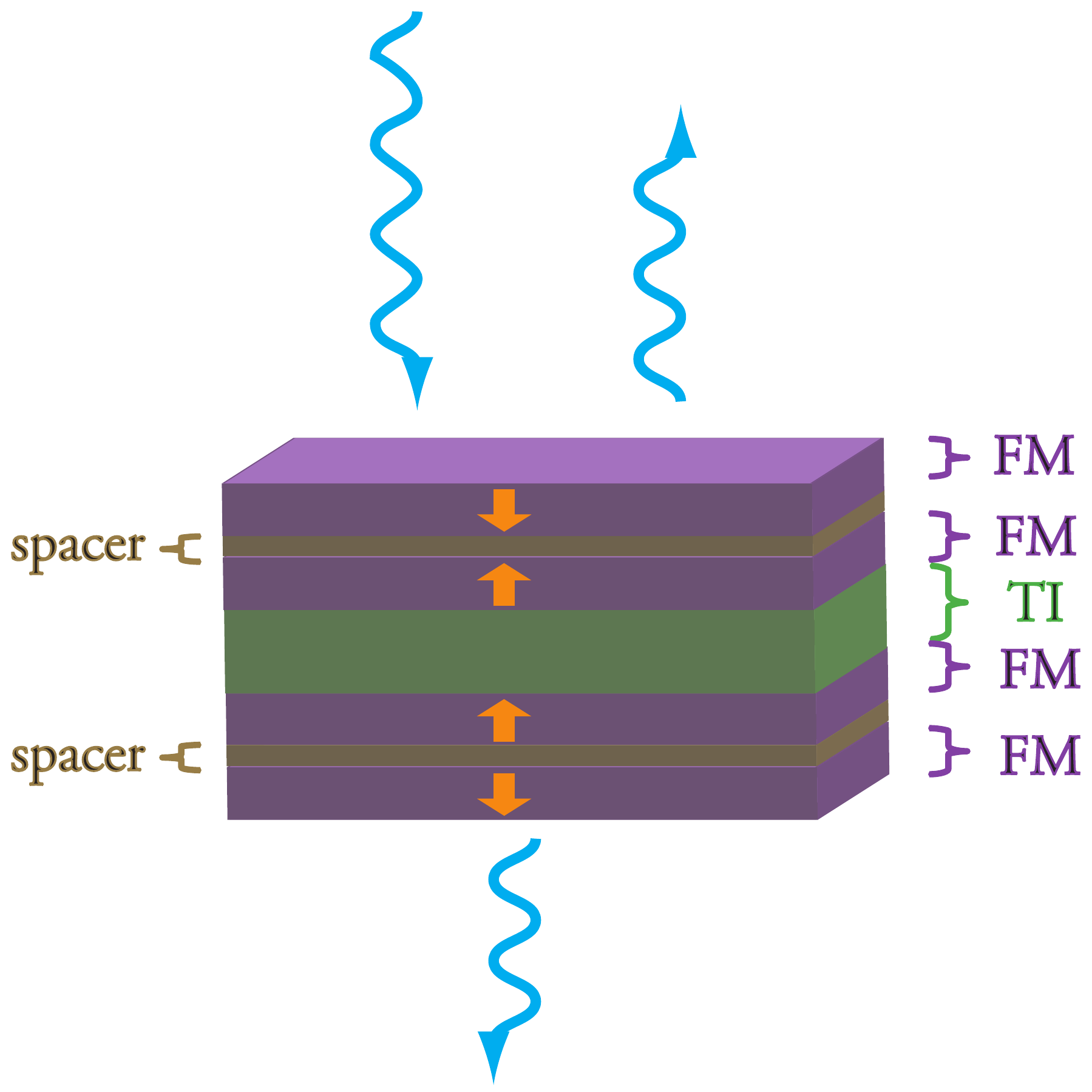} 
\caption{(Color online) Proposed experimental setup for observing the topological
  Kerr and Faraday effects. A topological insulator thin film is
  sandwiched between two sets of  ferromagnets (FM) with opposite
  magnetizations (indicated by straight arrows) stacked together,
  which induce an exchange field at the topological insulator
  surface. 
} \label{fig6}
\end{figure}
%
%

\textit{Breaking TRS in topological insulators} --- These properties of TIs reflect not intrinsically broken TRS breaking, but 
exceptional sensitivity to TRS-breaking perturbations. 
The most obvious physical source for the TRS-breaking coupling $\Delta$ is the Zeeman interaction
between spins and external magnetic fields, 
possibly produced by a permanent magnet layer under the TI thin film.
External fields, of course, also couple to the orbital degrees-of-freedom of the TI, and  
yield additional effects, which are not addressed here but are not necessarily unfavorable.
Another possibility, also mentioned in Ref.~[\onlinecite{SCZhangRev}], is to exploit exchange coupling 
to ferromagnetic thin films adjacent to the TI. (See
Fig.~\ref{fig6}).  In this case the ferromagnet would have to be insulating;
otherwise the topological insulator electronic structure would be altered by 
hybridization and the calculations presented here would not be directly relevant.
Proximity-induced exchange coupling 
between metallic states and ferromagnetic insulating GdN, for example, 
has been experimentally demonstrated
in a superconductor/insulating ferromagnet
multilayer thin-film structure \cite{GdN1}.
A key step in pursuing this strategy would be the identification of 
an insulating ferromagnet whose exchange coupling with Dirac-point TI states 
is in an interesting range; for the analysis presented here to be relevant the 
exchange coupling would have to be stronger than the TI surface state 
life-time broadening energy $\hbar/\tau$.  Of course, the ferromagnet layers themselves will
generically have a non-zero {\em AC} Hall conductivity, and will alter the 
magneto-optical response of the structure when present.  Because of their exceptional
magnetoelectric properties, the TI surface state response will tend to 
dominate if the ferromagnetic layers are thin.  The direct ferromagnet 
magneto-optical response can be mitigated by adding an 
additional set of ferromagnetic films with opposite magnetizations that sandwich the structure.
A related possibility is to surround the TI thin film not with ferromagnets but 
with antiferromagnetic layers,
similar to the exchange bias layers used in spin valves, which have
dominant spin orientations on the surface exposed to the TI.
Of all the topological insulators discovered to date, Bi$_2$Se$_3$ exhibits the largest bulk band gap. 
The predicted effects should therefore be most easily observable in Bi$_2$Se$_3$
since the low-frequency condition $\omega \ll E_{\mathrm{g}}$ can be more readily satisfied. 

This work was supported by the Welch Foundation and by the DOE under grant DE-FG03-02ER45985 .  The authors acknowledge 
helpful discussions with Jim Erskine, Andrew Essen, Byounghak Lee, Joel Moore, and Gennady Shvets.



\begin{thebibliography}{18}
\bibitem{half} We will address the precision of the half-quantized
  Hall effect at $\omega=0$ in a separate publication.
\end{thebibliography}
\end{document}